%% file: kn.tex
\documentclass[12pt]{article}

\usepackage{amsmath,amssymb,amsfonts,amsthm,amscd,bm}
\usepackage{graphicx}
\usepackage[T1]{fontenc}
\usepackage[utf8]{inputenc}
\usepackage[font=small,labelfont=bf,width=\textwidth]{caption}

\usepackage[a4paper,text={160mm,247mm},centering]{geometry}
\let\oldbfseries=\bfseries
\let\oldmdseries=\mdseries
\let\oldnormalfont=\normalfont
\renewcommand{\bfseries}{\oldbfseries\boldmath}
\renewcommand{\mdseries}{\oldmdseries\unboldmath}
\renewcommand{\normalfont}{\oldnormalfont\unboldmath}

\allowdisplaybreaks[3]
\input{macros}

\begin{document}

%\printlength\textwidth

\title{\bf \LARGE
The algebraic structure behind the  
derivative nonlinear Schr\" odinger equation 
}

\author{
G. S. Fran\c ca$^{ab\star}$, 
J. F. Gomes$^{b\dagger}$ and
A. H. Zimerman$^{b\ddagger}$
}

\date{}

\maketitle

\vspace{-2.5em}

\begin{center}
$^{a}${\it\small
Department of Physics, Cornell University, Ithaca, NY, USA
}

\vspace{0.5em}

$^{b}${\it\small 
Instituto de F\' isica Te\' orica---IFT/UNESP, S\~ ao Paulo, SP, Brazil
}
\end{center}

\vspace{1em}

%\title[The algebraic structure behind the DNLS equation]{The algebraic
%structure behind the derivative nonlinear Schr\" odinger equation}
%
%\author{G S Fran\c ca, J F Gomes and A H Zimerman}
%
%\ead{
%\mailto{guisf@ift.unesp.br},
%\mailto{jfg@ift.unesp.br},
%\mailto{zimerman@ift.unesp.br}
%}
%
%\address{
%Instituto de F\' isica Te\' orica - IFT/UNESP\\
%Rua Dr. Bento Teobaldo Ferraz, 271, Bloco II\\
%01140-070, S\~ ao Paulo - SP, Brazil
%}

\begin{abstract}
The Kaup-Newell (KN) hierarchy contains the derivative nonlinear
Schr\" odinger equation (DNLSE) amongst others interesting and important
nonlinear integrable equations. In this paper, a general higher grading
affine algebraic construction of integrable hierarchies is proposed and
the KN hierarchy is established  in terms of a $\hat{s\ell}_2$
Kac-Moody algebra and principal gradation. In this form, our spectral problem
is linear in the spectral parameter.
The positive and negative
flows are derived, showing that some interesting
physical models  arise from the same algebraic structure.  For instance,
the DNLSE is obtained as the second positive, while the
Mikhailov model as the first negative flows, respectively.
The equivalence between the latter and the massive Thirring model is
explicitly demonstrated also.
The algebraic dressing method is employed to construct
soliton solutions in a systematic manner for all members of the hierarchy.
Finally, the equivalence of the spectral problem introduced in this paper
with the usual one, which is quadratic in the spectral parameter, is
achieved by setting a particular automorphism of the affine algebra,
which maps the homogeneous into principal gradation.
%The construction proposed here avoids the need to
%introduce weight functions that were previously used in the revised
%form of the inverse scattering transform.
\end{abstract}

\vfill
{\it June/2013}

\vspace{-.7em}
\noindent\line(1,0){175}

{\footnotesize
$^{\star}$gsf47@cornell.edu, 
$^{\dagger}$jfg@ift.unesp.br, 
$^{\ddagger}$zimerman@ift.unesp.br
}

\newpage

\tableofcontents

\section{Introduction}
\label{sec:intro}

The \emph{derivative nonlinear Schr\" odinger equation} (DNLSE-I)
\begin{equation}\label{dnlse1}
i\pa_t \psi  + \pa_x^2 \psi \pm i \pa_x\(|\psi|^2 \psi\) = 0
\end{equation}
is a well known integrable model with interesting physical
applications. In particular, it describes nonlinear Alfv\' en waves
in plasma physics \cite{mio,mjolhus,mjolhus2,spangler,ruderman,fedun}
and the propagation of ultra-short pulses in nonlinear optics
\cite{tzoar,anderson_lisak,degasperis}.

Equation \eqref{dnlse1}, and also other related models that will be
mentioned in the following, has been extensively studied since a long
time ago. Its inverse scattering transform (IST) with a vanishing boundary
condition (VBC), $\psi\to0$ as $|x|\to\infty$,
was first solved in \cite{kaup_newell}. Equation \eqref{dnlse1} is one
of the nonlinear evolution equations comprising the Kaup-Newell (KN)
hierarchy.

The complete integrability of \eqref{dnlse1} and the hierarchy of
Hamiltonian structures of the KN hierarchy were constructed in 
\cite{gerdjikov_kulish}.
Moreover, the Riemann-Hilbert problem was considered as well as
expansions over the squared solutions \cite{gerdjikov_kulish}.
B\" acklund transformation for the KN hierarchy was obtained 
\cite{kundu_backlund}, and also a nonholonomic deformation of the hierarchy
was proposed \cite{kundu_deformation}.
A specific feature of these models is that they have a Lax operator
containing a \emph{quadratic} power on the spectral
parameter. More specifically, \eqref{dnlse1} and its related models
can be obtained from a zero-curvature representation where the standard
spectral problem is given by
\begin{equation}\label{kn_lax}
A_x = \begin{pmatrix}
\l^2 & \l q \\
\l r & -\l^2
\end{pmatrix}, \qquad \(\pa_x + A_x\)\Psi = 0,
\end{equation}
where $q=q(x,t)$ and $r=r(x,t)$ are the fields and $\l$ is the complex
spectral parameter.
Due to the quadratic power of $\l$ the original form of the
IST \cite{zakharov_shabat} has a divergent Cauchy
integral when $|\l|\to\infty$. This is the main reason for introducing
the revision in the IST \cite{kaup_newell}, through some weight functions
that control such divergence.
Let us recall that the Zakharov-Shabat approach \cite{zakharov_shabat}
was initially proposed to solve the Ablowitz-Kaup-Newell-Segur
(AKNS) spectral problem
\begin{equation}\label{akns_lax}
A_x = \begin{pmatrix}
\l & q \\
r & -\l
\end{pmatrix}.
\end{equation}
Another obvious difference worth mentioning is that in \eqref{kn_lax} the
fields are associated to $\l$, unlike \eqref{akns_lax}.
This implies in a \emph{higher grading} construction from the algebraic
formalism point of view, contrary to the AKNS construction
where the fields are in a \emph{zero grade} subspace.

For a nonvanishing boundary condition (NVBC),
$\psi\to\mbox{const.}$ as $|x|\to\infty$, an IST approach was also
proposed \cite{kawata_inoue,kawata_kobayashi_inoue}, but it is a
difficult procedure due to the appearance of a double-valued function of
$\l$ and, therefore, the IST had to be developed on
its Riemann sheets.
A more straightforward method introduces a convenient
affine parameter that avoids the construction of the Riemann
sheets \cite{chen_lam}. Recently, solutions with
VBC and also NVBC were constructed through Darboux/B\" acklund
transformations yielding explicit and useful formulas
\cite{steudel,xu_wang_darboux}. These works are all based on operator
\eqref{kn_lax} or the revised form of the IST \cite{kaup_newell}.

There are also two other known types of derivative nonlinear
Schr\" odinger equations, namely, the DNLSE-II \cite{chen_lee_liu}
\begin{equation}\label{dnlse2}
i\pa_t\psi + \pa_x^2\psi \mp 4i |\psi|^2\pa_x\psi = 0
\end{equation}
and the DNLSE-III \cite{gerdjikov}
\begin{equation}\label{dnlse3}
i\pa_t\psi + \pa_x^2\psi \pm 4i\psi^2\pa_x\psi^* + 8|\psi|^4\psi = 0.
\end{equation}
The gauge equivalence between \eqref{dnlse1}, \eqref{dnlse2} and
\eqref{dnlse3} was analyzed for the first time in \cite{gerdjikov} and
its Hamiltonian structures have also being extensively studied.
(Regarding gauge equivalent models see also \cite{kundu_gauge}.)
A hierarchy containing \eqref{dnlse2} within the Sato-Wilson dressing
formalism was also considered and it was shown that \eqref{dnlse2} can
be reduced to the fourth Painlev\' e equation \cite{kakei_kikuchi}.

The well known \emph{massive Thirring model}
\begin{equation}\label{thirr}
\begin{split}
i\pa_x v - mu + 2g |u|^2v &= 0, \\
i\pa_t u + mv - 2g |v|^2u &= 0,
\end{split}
\end{equation}
was proved to be integrable and solved through the IST for the first
time in \cite{mikhailov_thirring}
(see also \cite{orfanidis,thirring_kaup_newell}).
The relation between \eqref{thirr} and \eqref{dnlse1} was also
pointed out \cite{gerdjikov_kulish, gerdjikov}.
We will show this relation precisely through another model,
arising naturally from the first negative flow of the KN hierarchy, namely
\begin{equation}\label{modPLR}
\pa_x\pa_t\vphi - \vphi \mp 2i|\vphi|^2\pa_x \vphi = 0.
\end{equation}
This relativistically invariant model has attracted attention only 
recently although  it
was already proposed  a long time \cite{gerdjikov_kulish, gerdjikov}
and is known as the \emph{Mikhailov model}.
It is already known that \eqref{modPLR} is equivalent
to \eqref{thirr} \cite{gerdjikov_kulish,gerdjikov}.
Equation \eqref{modPLR} was also called Fokas-Lenells equation in recent works
and its multi-soliton solutions were obtained through Hirota's
bilinear method \cite{matsuno_fl}, where it was pointed out that they
have essentially the same form as those of \eqref{dnlse1} if one
introduces the potential $\psi=\pa_x\varphi$ and change the
dispersion relation. We will explain precisely the origin of this relation.
Equation \eqref{modPLR} was also referred as the
modified Pohlmeyer-Lund-Regge model \cite{kikuchi_mplr}
and it can be reduced to the third Painlev\' e equation.

It is important to mention that there are matrix or multi-field
generalizations of DNLS and Thirring like models
\cite{tsuchida_wadati,tsuchida_wadati2,tsuchida}.

The integrable properties of \eqref{dnlse1}, \eqref{dnlse2}, \eqref{dnlse3}
and \eqref{thirr}, like their soliton solutions and Hamiltonian hierarchies,
had already been thoroughly studied, specially in
\cite{kaup_newell,gerdjikov_kulish,gerdjikov,mikhailov_thirring}.
Nevertheless, these models, and more recently \eqref{modPLR}, continue
to attract attention and they have not being fully studied through
more recent techniques. One of the approaches to obtain soliton
solutions is through the dressing method proposed in
the pioneer paper \cite{zakharov_shabat}, which is connected to
the Riemann-Hilbert problem. Another approach occurs in
connection to $\tau$-functions and transformation groups
\cite{date_jimbo}, where solitons are obtained through vertex
operators. The precise relation between these two methods
were established in \cite{babelon_dressing} (see also
\cite{babelon_book} for a thoroughly explanation) and it enables one
to construct soliton solutions in a purely affine algebraic manner. Thus
we refer to this approach as the \emph{algebraic dressing method}.

Since then \cite{babelon_dressing}, there has been a significant
development of affine algebraic techniques
\cite{olive_vertex,gomes_soliton,aratyn_symmetry_flows,aratyn_construction}
relying on the algebraic structure underlying the
equations of motion, providing \emph{general} and \emph{systematic methods}.
This algebraic approach is well understood for models that fit into the
AKNS construction, where the fields are associated to \emph{zero grade}
operators. Nevertheless, there are some generalizations of this formalism,
for instance, the addition of fermionic fields to \emph{higher grading}
operators \cite{gervais_higher_grading,ferreira_affine_toda_matter} and
a generalization of the dressing method to include NVBC for the AKNS
construction \cite{dressing_nvc,mkdv_nvc,negative_even}.
Recently, it was proposed a higher grading construction
that includes the Wadati-Konno-Ichikawa hierarchy and the
algebraic dressing formalism was supplemented with reciprocal
transformations \cite{wki_spe}. Following this line of though,
it is important to embrace other known integrable models into such
algebraic formalism, to extend these techniques beyond the AKNS scheme.
This is accomplished here regarding the KN hierarchy. We introduce its
underlying algebraic structure and employ the algebraic
dressing method to construct
its soliton solutions systematically. Through gauge transformations we
also obtain solutions of the other related models like \eqref{dnlse2},
\eqref{dnlse3} and \eqref{thirr}, for instance. An important remark
is in order. We construct these models starting from a spectral problem
that is \emph{not quadratic} on the spectral parameter and it
simplifies the procedure to construct the solutions. We also prove
that our construction is equivalent to the quadratic one \eqref{kn_lax}.

Our work is thus organized as follows.
In section \ref{sec:general_hierarchy} we introduce a \emph{general}
higher grading affine algebraic construction of integrable hierarchies.
In section \ref{sec:kn_hierarchy}, when the algebra $\hat{A}_1$
with \emph{principal gradation} is chosen, this construction yields
the KN hierarchy as a particular case, where
\eqref{dnlse1} appears as the second positive
and \eqref{modPLR} as the first negative flows, respectively.
We introduce the explicit transformations relating equations \eqref{dnlse1},
\eqref{dnlse2} and \eqref{dnlse3}. Furthermore, we propose a transformation
that takes \eqref{modPLR} to the Thirring model \eqref{thirr},
demonstrating their equivalence.
In section \ref{sec:dressing} we employ the algebraic dressing method,
which along with representation theory of affine Lie algebras,
provides a systematic construction of the solutions for all
the models within the KN hierarchy. We also obtain the explicit solutions
of the other mentioned models.
In section \ref{sec:equivalency} we prove the equivalence between
our construction, which is linear in $\l$, and the usual one \eqref{kn_lax}.
Finally, section \ref{sec:conclusions} contains our concluding remarks.
We refer the reader to the appendix \ref{sec:algebra}
for the algebraic concepts involved in this paper.

\section{A general integrable hierarchy}
\label{sec:general_hierarchy}

Let $\alie$ be a semi-simple Kac-Moody algebra
with a grading operator $Q$.
The algebra is then decomposed into graded subspaces\footnote{See the
appendix \ref{sec:algebra} for the algebraic concepts.}
\begin{equation}
\alie = \sum_{m\in\integer}\alie^{(m)}, \qquad
\alie^{(m)} = \big\{ T_a^{n} \ | \ \big[Q,T_a^{n}\big]=mT_a^{n}\big\}.
\end{equation}
The parentheses in the superscript of an operator denote 
its grade according to $Q$ and should not be confused with the affine
index $n$ without parentheses, which is the power of the spectral parameter.

Let $\E{2}$ be a semi-simple element of \emph{grade two}. 
Define the kernel, $\ck$, and image, $\cm$, subspaces as follows
\begin{equation}\label{kernel_image}
\begin{split}
\ck &\equiv \big\{ T_a^{n} \in \alie \ | \ \big[ E^{(2)}, T_a^{n}\big] = 
0 \big\}, \\
\alie &= \ck \oplus \cm.
\end{split}
\end{equation}
From the Jacobi identity we conclude that
\begin{equation}\label{subset1}
\[\ck,\ck\]\subset\ck, \qquad \[\ck,\cm\]\subset\cm,
\end{equation}
and we assume the symmetric space
structure 
\begin{equation}\label{subset2}
\[\cm,\cm\]\subset\ck.
\end{equation}

Let $\F{1}[\bvphi]\in \cm^{(1)}$ be the operator containing 
the \emph{fields} of the models, i.e. choose all generators of \emph{grade one} 
in $\cm$,  $\cm^{(1)}=\blcurl R_1^{(1)}, \dotsc, R_r^{(1)} \brcurl$, and 
take the linear combination $\F{1}[\bvphi]=\vphi_1R_1^{(1)}+\dotsb+
\vphi_rR_r^{(1)}$ where $\bvphi=(\vphi_1,\dotsc,\vphi_r)$ and 
$\vphi_i=\vphi_i(x,t)$. 
Now we define the integrable hierarchy of PDE's starting from the zero 
curvature condition
\begin{equation}\label{zero_curvature}
\pa_xA_t - \pa_tA_x + \[A_x, A_t\] = 0
\end{equation}
with the potentials defined as follows
\begin{subequations}\label{potentials}
\begin{align}
A_x &\equiv \E{2} + \F{1}[\bvphi], \label{potential_u}\\
A_t &\equiv \sum_{m=-2M}^{2N}\D{m}[\bvphi], \label{potential_v}
\end{align}
\end{subequations}
where $N$ and $M$ are arbitrary fixed \emph{positive integers} that label
each model within the hierarchy and $\D{m}[\bvphi] \in \alie^{(m)}$.

Let us now show that with the algebraic structure defined by
\eqref{potentials}, 
the zero curvature equation \eqref{zero_curvature} can be solved 
nontrivially, yielding a PDE for each choice of $N$ and $M$.
Note that \eqref{potential_u} is well defined and independent of 
$N$ and $M$, therefore, we need to show that each $\D{m}$ 
in \eqref{potential_v} can be uniquely determined in terms of the fields.
Projecting \eqref{zero_curvature} into each graded subspace
we obtain the following set of equations
\begin{align}
\big[\E{2},\D{2N}\big] &=0 \qquad \text{grade $2N+2$}, \nnn
\big[\E{2},\D{2N-1}\big] + \big[\F{1},\D{2N}\big] &=0 
\qquad\text{grade $2N+1$}, \nnn
\pa_x\D{2N} + \big[\E{2},\D{2N-2}\big] + \big[\F{1},\D{2N-1}\big] &=0 \qquad 
\text{grade $2N$},\nnn
& \ \, \vdots \nnn
\pa_x\D{2} + \big[\E{2},\D{0}\big] + \big[\F{1},\D{1}\big] &=0 \qquad 
\text{grade $2$},\nnn
\pa_x\D{1} - \pa_t\F{1}+ \big[\E{2},\D{-1}\big] + \big[\F{1},\D{0}\big] &= 0 
\qquad \text{grade $1$}, \label{set_of_equations} \\
\pa_x\D{0} + \big[\E{2},\D{-2}\big] + \big[\F{1},\D{-1}\big] &= 0 
\qquad \text{grade 0},\nnn
& \ \, \vdots  \nnn
\pa_x\D{-2M+2} + \big[\E{2},\D{-2M}\big] + \big[\F{1},\D{-2M+1}\big] &=0
\qquad\text{grade $-2M+2$},\nnn
\pa_x\D{-2M+1} + \big[\F{1},\D{-2M}\big] &= 0 \qquad \text{grade $-2M+1$},\nnn
\pa_x\D{-2M} &= 0 \qquad \text{grade $-2M$}. \nn
\end{align}
These equations can be solved recursively, starting from grade $2N+2$ 
until grade $2$ and from grade $-2M$ until grade zero. 
Each equation still splits into $\ck$ and $\cm$ components 
according to \eqref{subset1} and 
\eqref{subset2}. We also have the $\ck$ component
of the grade one equation as a constraint. 
Thus each $\D{m}$ is determined in terms of the fields $\bvphi$ and, 
consequently, the equations of motion are obtained from the $\cm$ component 
of the grade one projection 
\begin{equation}\label{movement}
\pa_x\D{1}_\cm-\pa_t \F{1} + \[ \E{2}, \D{-1}_\cm \] +
\[ \F{1}, \D{0}_\ck \] = 0.
\end{equation}

The potential \eqref{potential_v} generates \emph{mixed flows} 
\cite{gomes_mixed}. If one is interested in \emph{positive flows} only, 
yielding the positive part of the hierarchy, one must 
restrict the sum as 
\begin{equation}\label{positive_flows}
A_t\equiv\sum_{m=1}^{2N}\D{m}[\bvphi]
\end{equation}
and the equations of motion then simplify to
\begin{equation}\label{motion_2}
\pa_x\D{1}_\cm-\pa_t \F{1} = 0.
\end{equation}
For \emph{negative flows} one must restrict the sum as
\begin{equation}\label{negative_flows}
A_t\equiv\sum_{m=0}^{-2N}\D{m}[\bvphi]
\end{equation}
and the equations of motion are then given by
\begin{equation}\label{movement3}
\pa_t\F{1}-\[ \E{2}, \D{-1}_\cm\]-\[ \F{1},\D{0}_\ck\]=0.
\end{equation}

Therefore, we have shown that the hierarchy defined by 
\eqref{potentials} solves the zero curvature equation \eqref{zero_curvature}. 
This construction is valid for an arbitrary graded affine Lie 
algebra $\alie$ and the hierarchy is defined 
through the choice of $\langle \alie,Q,\E{2}\rangle$, leading to an immediate 
algebraic classification. 
Each choice of positive integers $N$ and $M$ yields one mixed model.
For positive or negative flows the models are labeled by $N$ only.

\section{The Kaup-Newell hierarchy}
\label{sec:kn_hierarchy}

Let us take the previous construction with the loop-algebra
$\llie=\tilde{A}_1=\{\H{n},\Ea{n},\Eb{n}\}$ and
\emph{principal gradation} $Q=\tfrac{1}{2}\H{0}+2\hd$, leading to the
decomposition $\alie^{(2m)}=\{\H{m}\}$ and 
$\alie^{(2m+1)}=\{\Ea{m}, \Eb{m+1}\}$. The semi-simple 
element is chosen as $E^{(2)}=\H{1}$ and then \eqref{kernel_image} is given
by
\begin{equation}\label{graded_sub}
\ck^{(2m)}=\big\{\H{m} \big\}, \qquad
\cm^{(2m+1)}=\big\{\Ea{m}, \Eb{m+1}\big\}.
\end{equation}
The operator containing the fields must now have the form 
$F^{(1)} = q(x,t)\Ea{0} + r(x,t)\Eb{1}$
and \eqref{potentials} then reads
\begin{subequations}\label{kn_potentials}
\begin{align}
A_x &= \H{1} + q\Ea{0} + r\Eb{1}, \label{kn_p1} \\
A_t &= \sum_{m=1}^{2N}\D{m} 
\qquad \text{or} \qquad
A_t=\sum_{m=0}^{-2N}\D{m}, \label{kn_p2}
\end{align}
\end{subequations}
where $\D{2m}=c_{2m}\H{m}$ and $\D{2m+1}=a_{2m+1}\Ea{m}+b_{2m+1}\Eb{m+1}$. 
The coefficients $a_{2m+1}$, $b_{2m+1}$ and $c_{2m}$ will be determined 
in terms of the fields $q$ and $r$ by solving the zero curvature equation,
as explained previously in \eqref{set_of_equations}.
The first equality in \eqref{kn_p2} is valid for the \emph{positive flows}, 
while the second one for the \emph{negative flows}.

\paragraph{Comment.} If instead of the principal gradation one considers
the \emph{homogeneous} gradation $Q=\hd$, yielding
$\alie^{(m)}=\lcurl \Ea{m}, \Eb{m}, \H{m}\rcurl$, and the semi-simple element  
$\E{2}=\H{2}$, we obtain the following Lax pair for the KN hierarchy
\begin{subequations}\label{standard_lax_kn}
\begin{align}
A_x &= \H{2} + q\Ea{1} + r\Eb{1}, \label{standard1} \\
A_t &= \sum_{m=1}^{2N}\D{m} \qquad \mbox{or} \qquad
A_t = \sum_{m=0}^{-2N}\D{m},
\end{align}
\end{subequations}
where $\D{m} = a_m\Ea{m} + b_m\Eb{m} + c_m\H{m}$.
The operator \eqref{standard1} is exactly the standard 
one found in the literature \eqref{kn_lax} \cite{kaup_newell,
gerdjikov_kulish,gerdjikov,mikhailov_thirring} (see the matrix representation
in appendix \ref{sec:algebra}). With the construction
\eqref{standard_lax_kn} we obtain precisely the same equations of motion as
the construction \eqref{kn_potentials}, which will be derived in the
sequel. Moreover, we will demonstrate the equivalence between
both constructions in section \ref{sec:equivalency}. 
Note, however, that \eqref{kn_p1} \emph{does not have a quadratic power} 
on the spectral parameter, contrary to \eqref{standard1}.
The convenience of using \eqref{kn_p1} appears clearly when constructing the 
solutions of the models within the hierarchy.

\subsection{Positive flows}
The models within the positive part of the hierarchy 
are obtained from the zero curvature equation
\begin{equation}\label{kn_positive}
\[\pa_x+\H{1}+q\Ea{0}+r\Eb{1}, 
\pa_t+\D{2N}+\D{2N-1}+\dotsb+\D{1}\]=0.
\end{equation}
For $N=1$ we have the trivial equations 
$\pa_t q=\pa_xq$ and $\pa_t r = \pa_x r$. For $N=2$, after solving
each equation in \eqref{set_of_equations}, we get the following
equations of motion
\begin{subequations}\label{dnls}
\begin{align}
2\pa_t q + \pa_x^2q + \pa_x\(q^2r\) &= 0, \\
2\pa_t r - \pa_x^2r + \pa_x\( qr^2\) &=0,
\end{align}
\end{subequations}
whose explicit Lax pair is given by
\begin{subequations}\label{dnls_lax}
\begin{align}
A_x&=\H{1}+q\Ea{0}+r\Eb{1}, \\
A_t&=\H{2}+q\Ea{1}+r\Eb{2}-\tfrac{1}{2}qr\H{1}
- \tfrac{1}{2}\( q^2r+\pa_xq\)\Ea{0}-
\tfrac{1}{2}\( qr^2-\pa_xr\)\Eb{1}.
\end{align}
\end{subequations}
Taking \eqref{dnls} under the transformations 
$x \to ix$, $t \to 2it$, $q = \psi$ and 
$r = \pm \psi^*$, we obtain precisely equation \eqref{dnlse1},
\begin{equation}\label{dnlse}
i\pa_t \psi + \pa_x^2 \psi \pm i\pa_x\(|\psi|^2\psi\) = 0.
\end{equation}
Considering $N=3$, and after solving \eqref{set_of_equations}, we obtain 
the model
\begin{subequations}\label{kn3}
\begin{align}
4\pa_t q - \pa_x^3 q - 3\pa_x\(qr\pa_xq\) - \tfrac{3}{2}\pa_x\(q^3r^2\) &= 0,\\
4\pa_t r - \pa_x^3 r + 3\pa_x\(qr\pa_xr\) - \tfrac{3}{2}\pa_x\(q^2r^3\) &= 0,
\end{align}
\end{subequations}
together with its Lax pair
\begin{subequations}
\begin{align}
A_x &= \H{1} + q\Ea{0} + r\Eb{1}, \\
A_t &= \H{3} + q\Ea{2}+r\Eb{3} - \tfrac{1}{2}qr\H{2}
-\tfrac{1}{2}\( q^2r+\pa_xq\)\Ea{1} 
\nn \\ & \qquad
-\tfrac{1}{2}\( qr^2 - \pa_xr\)\Eb{2} 
+\tfrac{1}{4}\( r\pa_xq - q\pa_xr + \tfrac{3}{2}q^2r^2\)\H{1}
\nn \\ & \qquad
+\tfrac{1}{4}\( \pa_x^2q + 3qr\pa_xq + \tfrac{3}{2}q^3r^2\) \Ea{0} 
+\tfrac{1}{8}\( \pa_x^2r - 3qr\pa_xr + \tfrac{3}{2}q^2r^3 \) \Eb{1}.
\end{align}
\end{subequations}
The system \eqref{kn3} under  
$x\to ix$, $t\to -4it$, $q=\psi$ and $r=\pm \psi^*$ becomes
\begin{equation}
\pa_t\psi - \pa_x^3\psi \mp 3i \pa_x\(|\psi|^2\pa_x\psi\)
+\tfrac{3}{2}\pa_x\(|\psi|^4 \psi \) = 0.
\end{equation}
%The equations \eqref{dnls} and \eqref{kn3} are well known \cite{kaup_newell}.
Continuing in this way for $N=4,5,\dotsc$ one can 
obtain higher order nonlinear PDE's.

\subsection{Negative flows} 
The negative flows of the KN hierarchy are obtained from the zero curvature 
equation
\begin{equation}\label{kn_negative}
\[\pa_x+\H{1}+q\Ea{0}+r\Eb{1}, 
\pa_t+\D{-2N}+\D{-2N+1}+\dotsb+\D{0}\]=0.
\end{equation}
Taking $N=1$, after solving \eqref{set_of_equations}, we obtain the 
following nonlocal field equations
\begin{subequations}\label{mpl_nloc}
\begin{align}
\tfrac{1}{4}\pa_t q - \int_{-\infty}^x qdx' + 
q\int_{-\infty}^x qdx' \int_{-\infty}^x rdx' &= 0, \\
\tfrac{1}{4}\pa_t r - \int_{-\infty}^x rdx' - 
r\int_{-\infty}^x qdx'\int_{-\infty}^x rdx' &=0.
\end{align}
\end{subequations}
These equations can be cast in a local form if we introduce new 
fields defined by
\begin{equation}\label{fields_derivative}
q \equiv \pa_x \phi, \qquad
r \equiv \pa_x \rho,
\end{equation}
and then we obtain the relativistically invariant
\emph{Mikhailov model} \cite{gerdjikov_kulish,gerdjikov}
\begin{subequations}\label{mpl}
\begin{align}
\tfrac{1}{4}\pa_x\pa_t \phi - \phi + \phi\rho \pa_x\phi & = 0, \\
\tfrac{1}{4}\pa_x\pa_t \rho - \rho - \phi\rho \pa_x\rho & = 0 .
\end{align}
\end{subequations}
Its explicit Lax pair is given by
\begin{subequations}\label{lax_mpl}
\begin{align}
A_x &= \H{1}+\pa_x\phi_1 \Ea{1}+\pa_x\rho\Eb{1},\\
A_t &= \H{-1}+2\phi\Ea{-1}-2\rho\Eb{0}+2\phi\rho\H{0}.
\end{align}
\end{subequations}
Taking \eqref{mpl} with $x \to \tfrac{i}{2}x$, $t \to -\tfrac{i}{2}t$, 
$\phi=\vphi$ and $\rho=\pm \vphi^*$ we obtain exactly \eqref{modPLR},
\begin{equation}\label{mplr}
\pa_x\pa_t \vphi - \vphi \mp 2i |\vphi|^2\pa_x \vphi = 0.
\end{equation}
This model can also be derived from the 
following Lagrangian
\begin{equation}
\mathcal{L} = \tfrac{1}{2}\(\pa_x\vphi\pa_t\vphi^* + 
\pa_x\vphi^*\pa_t \vphi\) + |\vphi|^2
\mp \tfrac{i}{2}|\vphi|^2\(\vphi\pa_x\vphi^* - \vphi^*\pa_x\vphi\).
\end{equation}
If one consider $N=2,3,\dotsc$ it is possible
to obtain higher order integro-differential equations like \eqref{mpl_nloc}.

\subsection{Relations between DNLS type equations}
Let us now introduce the transformations connecting the three types of
DNLS equations.
From \eqref{dnls} it immediately  follows the continuity equation
\begin{equation}\label{cont_dnls}
\pa_t\(qr\) + \pa_x j = 0, \qquad
j = \tfrac{1}{2}\(r\pa_xq - q\pa_xr\) + \tfrac{3}{4}\(qr\)^2.
\end{equation}
Let us define new fields through the following gauge transformation%
\footnote{It will be shown that, in fact, $\mathcal{J}$ is a local function.}
\begin{equation}\label{tilde_fields}
\tq \equiv -\tfrac{1}{2}qe^{c \mathcal{J}}, \qquad
\tr \equiv \tfrac{1}{2}re^{-c \mathcal{J}}, \qquad
\mathcal{J} \equiv \int_{-\infty}^x qrdx',
\end{equation}
where $c$ is a constant. Upon using \eqref{cont_dnls} the equations of
motion \eqref{dnls} can be written in terms of these new fields, yielding
\begin{subequations}\label{tilde_eqs}
\begin{align}
2\pa_t\tq + \pa_x^2\tq - A\tq^3\tr^2 
+ B\tq^2\pa_x\tr - C\pa_x\(\tq^2\tr\) &= 0, \\
2\pa_t\tr - \pa_x^2\tr + A\tq^2\tr^3 
+ B\tr^2\pa_x\tq - C\pa_x\(\tq\tr^2\) &=0,
\end{align}
\end{subequations}
where $A=8c\(2c-1\)$, $B=4c$ and $C=4\(1-c\)$. Note that in 
\eqref{tilde_eqs} we have a fifth order nonlinearity and two types of 
derivative nonlinear terms. Let us use the same transformation 
leading to equation
\eqref{dnlse}, i.e. $x\to ix$, $t\to 2it$, $\tq=\tpsi$ 
and $\tr=\pm\tpsi^*$. If besides this we set $c = \tfrac{1}{2}$ we obtain 
the equation \eqref{dnlse2},
\begin{equation}\label{dnls2}
i\pa_t\tpsi + \pa_x^2\tpsi \mp 4 i |\tpsi|^2\pa_x\tpsi = 0.
\end{equation}
On the other hand, if we set $c = 1$ we obtain the equation \eqref{dnlse3},
\begin{equation}\label{dnls3}
i\pa_t\tpsi + \pa_x^2\tpsi \pm 4 i \tpsi^2\pa_x\tpsi^* + 8|\tpsi|^4\tpsi = 0.
\end{equation}
Therefore, the transformation \eqref{tilde_fields} connects explicitly
the three types of DNLS equations. If one knows a solution of
\eqref{dnls} it is possible to obtain a solution of \eqref{tilde_eqs}, 
which in particular yields solutions of \eqref{dnls2} and \eqref{dnls3}.

\subsection{The massive Thirring model}
The massive Thirring model is obtained from the 
Lagrangian
\begin{equation}\label{thirring_lagrangian}
\mathcal{L} = \bar{\Phi}\(i \gamma^\mu\pa_\mu - m\) \Phi 
+\dfrac{g}{2}J_\mu J^{\mu}, \qquad J^\mu = \bar{\Phi}\gamma^\mu\Phi,
\end{equation}
where $m$ is the mass, $g$ is the coupling constant and
\begin{equation}
\Phi = \begin{pmatrix} u \\ v \end{pmatrix}, \qquad
\gamma^0 = \gamma_0 = \sigma_1 = \begin{pmatrix} 0 & 1 \\ 1 & 0 \end{pmatrix}, 
\qquad
\gamma^1 = -\gamma_1 = i\sigma_2 = \begin{pmatrix} 0 & 1 \\ -1 & 0 
\end{pmatrix}.
\end{equation}
The $\sigma_i$ are the usual Pauli matrices and
$\bar{\Phi} \equiv \Phi^{\dagger}\gamma^0$.
The equations of motion are then given by 
\begin{equation}
i\gamma^\mu\pa_\mu \Phi - m\Phi + gJ_\mu\gamma^\mu \Phi = 0,
\end{equation}
or written in component form and in the light cone coordinates
$x \equiv \tfrac{1}{2}\(x^1+x^0\)$ and $t\equiv\tfrac{1}{2}\(x^1-x^0\)$,
we thus have
\begin{subequations}\label{thirr1}
\begin{align}
i\pa_x v - m u + 2g |u|^2 v &= 0, \\
i\pa_t u + m v - 2g |v|^2 u &= 0.
\end{align}
\end{subequations}
%The IST for this  model was solved in \cite{thirring_kaup_newell}
%and its spectral problem has essentially the same form as the 
%DNLSE-I \eqref{dnlse1} spectral problem. 
An important conclusion pointed out in \cite{thirring_kaup_newell} is that 
the solutions of \eqref{thirr1} cannot, in general, correspond
to solutions of the sine-Gordon model. However, as will be shown below,
its solutions can be obtained, in general, from the solutions of 
the model \eqref{mplr}.

Consider \eqref{mpl} under the transformations
$x \to - \tfrac{i}{2} x$, $t \to \tfrac{im^2}{2} t$ and 
$\phi = \vphi = \rho^*$. Thus we have the equation of motion given by
\begin{equation}\label{mplr2}
\pa_x\pa_t \vphi = m^2 \vphi - 2i m^2 |\vphi|^2\pa_x\vphi,
\end{equation}
from which it follows the continuity equation
\begin{equation}\label{conservation}
\pa_t|\pa_x \vphi|^2  = m^2 \pa_x|\vphi|^2.
\end{equation}
Let us define new fields through the following relations
\begin{equation}\label{new_fields}
u \equiv \dfrac{1}{\sqrt{2g}} \( \pa_x \vphi\) e^{i \mathcal{J}}, \qquad
v \equiv -\dfrac{im}{\sqrt{2g}} \vphi e^{i \mathcal{J}}, \qquad
\mathcal{J} \equiv \int_{-\infty}^x|\pa_{x'} \vphi|^2dx'.
\end{equation}
Then, calculating $\pa_x v$ and also $\pa_t u$, making use 
of \eqref{mplr2} and \eqref{conservation}, after
writing the result in terms of the fields $u$ and $v$ we get precisely
the equations \eqref{thirr1}. Therefore, if $\varphi$ is a solution of 
\eqref{mplr2}, then \eqref{new_fields} yields a solution of the Thirring 
model \eqref{thirr1}\footnote{%
If we consider the same kind of transformation for both equations
\eqref{mpl}, without requiring $\rho=\phi^*$, and define
$\chi_1 \equiv \tfrac{1}{\sqrt{2g}}\pa_x\phi e^{i\mathcal{J}}$,
$\chi_2 \equiv \tfrac{1}{\sqrt{2g}}\pa_x\rho e^{-i\mathcal{J}}$,
$\chi_3 \equiv \tfrac{im}{\sqrt{2g}}\rho e^{-i\mathcal{J}}$ and
$\chi_4 \equiv -\tfrac{im}{\sqrt{2g}}\phi e^{i\mathcal{J}}$,
where $\mathcal{J}\equiv \int_{-\infty}^x \pa_{x'}\phi\pa_{x'}\rho dx'$,
we obtain the four component Thirring like model considered
in \cite{tsuchida}, equation $(2.31)$.}.

\section{Dressing approach to the KN hierarchy}
\label{sec:dressing}

We now employ the algebraic dressing method  \cite{babelon_dressing}
to construct soliton solutions for the KN hierarchy, assuming
$q\to0$ and $r\to0$ when $|x|\to\infty$.
To extract the fields in the dressing formalism it is necessary to employ a
highest weight representation of the algebra, so we need 
the full Kac-Moody algebra $\hat{A}_1$ including the central term.
Thus, consider the Lax pair \eqref{kn_potentials} but  
with $A_x$ in the following slightly different form
\begin{equation}\label{lax_dress}
A_x = \H{1} + q\Ea{0} + r\Eb{1} - \(\pa_x\nu - 2t\d_{N+1}\)\hc, 
\end{equation}
where $\nu$ is a function to be determined. Note that the 
central term does not change the equations of motion,
since it commutes with all other generators. 
The vacuum solution, obtained by setting 
$q\to 0$, $r\to 0$ and $\nu \to 0$, is then given by
\begin{subequations}\label{vacuum}
\begin{align}
\bar{A}_x &=  \H{1} + 2t\delta_{N+1,0}\hc, \label{vac1}\\
\bar{A}_t &= \H{N}, \label{vac2} 
\end{align}
\end{subequations}
where $N=2,3,\dotsc$ for the \emph{positive flows} and $N=-1,-2,\dotsc$ for
the \emph{negative flows}. Note that \eqref{vac1} remains the same for 
every model, unless for the central term, while \eqref{vac2} changes 
according to each model. 
The potentials \eqref{vacuum} 
can still be written in the pure gauge form 
$\bar{A}_\mu = -\pa_\mu\bar{\Psi}\bar{\Psi}^{-1}$ with
\begin{equation}\label{psi_vacuum}
\bar{\Psi} = e^{-\H{1}x}e^{-\H{N}t}e^{-2xt\d_{N+1,0}\hc}.
\end{equation}
The idea of the dressing method is to obtain the
general potentials $A_\mu$, with a nontrivial field configuration, out 
from the vacuum $\bar{A}_\mu$, through the gauge transformations
\begin{equation}\label{gauge}
A_\mu = \Thpm \bar{A}_\mu \Thpm^{-1} - \pa_\mu\Thpm \Thpm^{-1}.
\end{equation}
Furthermore, we have the gauge freedom $\Psi \to \Psi'=\Psi g$
where $g$ is a constant group element. Hence, the dressing operators must 
satisfy $\Thp\bar{\Psi} = \Thm\bar{\Psi}g$, which is 
the Riemann-Hilbert problem
\begin{equation}\label{rh}
\Thm^{-1}\Thp = \bar{\Psi} g \bar{\Psi}^{-1}.
\end{equation}
Assuming a Gauss decomposition, the dressing operators can be
further factorized as
\begin{subequations}\label{gauss}
\begin{align}
\Thp &= e^{\A{0}}e^{\B{1}}e^{\B{2}}\dotsc \\
\Thm &= e^{\B{0}}e^{\B{-1}}e^{\B{-2}}\dotsc
\end{align}
\end{subequations}
where $\A{0}$ and $\B{m}$ are graded elements.
According to the principal gradation,
these operators must have the following form
\begin{equation}\label{Bm}
\B{2m+1} = \chi_{2m+1}\Ea{m}+\psi_{2m+1}\Eb{m+1}, \qquad
\B{2m} = \phi_{2m}\H{m},
\end{equation}
where the fields $\chi_{2m+1}$, $\psi_{2m+1}$ and $\phi_{2m}$ 
are now our unknowns.

It is enough to consider \eqref{gauge} for the Lax operator $A_x$. 
Taking it first with the operator $\Thp$, the projection into 
the \emph{grade zero} subspace yields
\begin{equation}\label{A0}
\A{0} = \nu\hc.
\end{equation}
The projection into the \emph{grade one} subspace yields
$q\Ea{0}+r\Eb{1} = -\pa_x\B{1}$, therefore
\begin{equation}\label{B1}
q = -\pa_x\chi_1, \qquad r = -\pa_x\psi_1.
\end{equation}
In this way, by considering higher grade projections we can 
determine all operators appearing in $\Thp$, 
but just relations \eqref{B1} are enough for our purposes. 
Now let us consider \eqref{gauge} with the operator $\Thm$. Taking
the \emph{grade two} projection we have $\[ \H{1}, \B{0} \] = 0$, therefore 
\begin{equation}\label{B0}
\B{0} = \phi_0\H{0}.
\end{equation}
Note that we already have a central term in \eqref{A0} so we do not need
to include another one in $\B{0}$. The field $\phi_0$ will be determined
by the next lower grades. Considering the \emph{grade one} 
projection we obtain
\begin{equation}\label{B-1}
q = -2\chi_{-1}e^{2\phi_0}, \qquad r = 2\psi_{-1}e^{-2\phi_0}.
\end{equation}
The \emph{grade zero} projection gives one equation for $\phi_{-2}$, which we
do not need, and also an equation for $\phi_0$ which is then given by
\begin{equation}\label{phi0}
\phi_0 = -\tfrac{1}{2}\int_{-\infty}^x qrdx'.
\end{equation}

Let us point out some important facts that came out naturally from these
approach. 
%Firstly, keep in mind that the fields 
%$\chi_j$, $\psi_j$ and $\phi_j$ are directly related to the 
%tau functions, as will be shown shortly, so they are considered to be
%the solutions of the models.
Comparing \eqref{B1} with \eqref{fields_derivative} we
see that the solutions of the first negative flow \eqref{mpl} are given by
\begin{equation}\label{sol_neg}
\phi = -\chi_{1}, \qquad \rho = -\psi_{1},
\end{equation}
while the solutions of \eqref{dnls} contains an extra derivative \eqref{B1}.
This explains why the solutions of \eqref{dnlse} are connected to those 
of \eqref{mplr} through the potential variable $\psi=\pa_x\vphi$ 
\cite{matsuno_fl,tsuchida}. Comparing \eqref{B-1} and \eqref{phi0} with
the gauge transformation \eqref{tilde_fields}, we note that the form of 
the transformation connecting the three types of DNLS equations are 
already contained in the dressing operators. The same also applies
to the relations \eqref{new_fields}. Moreover, we see from 
\eqref{phi0} that $\mathcal{J}=-2\phi_0$.
Precisely for the case $c=1$, corresponding 
to equation \eqref{dnls3}, we have from \eqref{B-1} that
$\tq = \chi_{-1}$ and $\tr = \psi_{-1}$. For 
equations \eqref{tilde_eqs} and \eqref{thirr1} we need to 
include the arbitrary constants. The term $e^{\pm i\phi_0}$ is precisely
the weight function introduced in the revised form of the 
IST \cite{kaup_newell}.

\subsection{Tau functions}
We now introduce an important class of functions that contain
the explicit space-time dependence of the solutions. They are directly
related to the fields $\chi_{2m+1}$, $\psi_{2m+1}$ and $\phi_{2m}$ 
of \eqref{Bm} and, consequently, to the physical fields in 
the equations of motion through the relations \eqref{B1}--\eqref{sol_neg}.
Consider the highest weight states\footnote{See the appendix
\ref{sec:algebra}.} 
$\{\ket{\l_0}, \ket{\l_1}\}$ and let us also introduce the following
\emph{notation} for convenience\footnote{These are not highest weight
states of the algebra.}
\begin{equation}
\ket{\l_2} \equiv \Ea{-1}\ket{\l_0}, \qquad
\ket{\l_3} \equiv \Eb{0}\ket{\l_1}.
\end{equation}
The procedure to extract the fields in the dressing approach
is to project the left hand side of \eqref{rh} between appropriate
states. Thus we have
\begin{equation}\label{chi_fields}
\begin{split}
e^{\nu} &= \bra{\l_0} \Thm^{-1}\Thp \ket{\l_0}, \\
\psi_1e^{\nu} &= \bra{\l_0} \Thm^{-1}\Thp \ket{\l_2}, \\
\chi_{-1}e^{\nu} &= -\bra{\l_2} \Thm^{-1}\Thp \ket{\l_0}, 
\end{split}\qquad
\begin{split}
e^{\nu-\phi_0} &= \bra{\l_1} \Thm^{-1}\Thp \ket{\l_1}, \\
\chi_1e^{\nu-\phi_0} &= \bra{\l_1} \Thm^{-1}\Thp \ket{\l_3}, \\
\psi_{-1}e^{\nu-\phi_0} &= -\bra{\l_3} \Thm^{-1}\Thp \ket{\l_1}.
\end{split}
\end{equation}
Note that the right hand side of \eqref{rh} contains
the explicit space-time dependence through \eqref{psi_vacuum}, so we define 
the $\tau$-\emph{functions} as
\begin{equation}\label{tau}
\tau_{ab} \equiv \bra{\l_a} \bar{\Psi}g\bar{\Psi}^{-1} \ket{\l_b},
\end{equation}
where $a,b=0,1,2,3.$
The $\tau$-functions are classified according to the arbitrary group 
element $g$. Combining the results of \eqref{chi_fields} 
and \eqref{tau} we have
\begin{equation}\label{tau_ansatz}
\phi_0 = \ln\(\dfrac{\tau_{00}}{\tau_{11}}\), \quad
\psi_1 = \dfrac{\tau_{02}}{\tau_{00}}, \quad
\chi_1 = \dfrac{\tau_{13}}{\tau_{11}}, \quad
\psi_{-1} = -\dfrac{\tau_{31}}{\tau_{11}}, \quad
\chi_{-1} = -\dfrac{\tau_{20}}{\tau_{00}}.
\end{equation}
From these relations we can express the solutions of all previous
models in terms of $\tau$-functions, which can be algebraically
calculated if we have an appropriate form for the group element $g$. 
For instance, from \eqref{sol_neg} the solutions
of \eqref{mpl} are expressed as
\begin{equation}\label{sol_mpl}
\phi = -\dfrac{\tau_{13}}{\tau_{11}}, \qquad
\rho = -\dfrac{\tau_{02}}{\tau_{00}},
\end{equation}
while from \eqref{B1} we have the solutions for the positive flows, 
like \eqref{dnls} and \eqref{kn3}, given by
\begin{equation}\label{sol_pos}
q = -\pa_x\(\dfrac{\tau_{13}}{\tau_{11}}\), \qquad
r = -\pa_x\(\dfrac{\tau_{02}}{\tau_{00}}\). 
\end{equation}
The integral appearing in the gauge transformations 
\eqref{tilde_fields} can be obtained from \eqref{phi0} yielding
\begin{equation}\label{integral}
\mathcal{J}= 2\ln\dfrac{\tau_{11}}{\tau_{00}},
\end{equation}
showing that it is indeed a local function.
Then, the solutions of \eqref{tilde_eqs} are given by
\begin{equation}\label{sol_tilde}
\tq = \dfrac{1}{2}
\(\dfrac{\tau_{11}}{\tau_{00}}\)^{2c}
\pa_x\(\dfrac{\tau_{13}}{\tau_{11}}\), \qquad
\tr = -\dfrac{1}{2}
\(\dfrac{\tau_{00}}{\tau_{11}}\)^{2c}
\pa_x\(\dfrac{\tau_{02}}{\tau_{00}}\).
\end{equation}
In particular, for equation \eqref{dnls3} ($c=1$) it is easier
to use directly \eqref{B-1} thus
\begin{equation}\label{sol_dnls3}
\tq = -\dfrac{\tau_{20}}{\tau_{00}}, \qquad
\tr = -\dfrac{\tau_{31}}{\tau_{11}}.
\end{equation}
Recall that we must further impose the transformations
$x\to ix$, $t\to 2it$, and $\tau_{31}/\tau_{11}=\pm\(\tau_{20}/\tau_{00}\)^*$.
For the massive Thirring model \eqref{thirr1} we obtain 
from \eqref{new_fields} the following solution
\begin{equation}\label{sol_thirr}
u = -\dfrac{1}{\sqrt{2g}}
\dfrac{\tau_{11}}{\tau_{00}}
\pa_x\(\dfrac{\tau_{13}}{\tau_{11}}\), \qquad
v = \dfrac{im}{\sqrt{2g}}\dfrac{\tau_{13}}{\tau_{00}}.
\end{equation}
We must further impose the transformations $x\to -\tfrac{i}{2} x$ and
$t\to \tfrac{im^2}{2} t$ in the space-time dependence of the $\tau$-functions
and also $\tau_{02}/\tau_{00}=\(\tau_{13}/\tau_{11}\)^*$. 

\subsection{Vertex operators}
In order to be able to evaluate \eqref{tau} explicitly let us 
assume that $g$ is given in the following form \cite{babelon_dressing,
olive_vertex}
\begin{equation}\label{vertex}
g \equiv \prod_{j=1}^{n}\exp\(\Gamma_j\), \qquad \Gamma_j = \Gamma\(\k_j\)
\end{equation}
where $\Gamma_j$ is a \emph{vertex operator} depending
on a \emph{complex} parameter $\kappa_j$. Furthermore, let us assume that
the vertex operator satisfy an eigenvalue commutation 
relation with \eqref{vacuum},
\begin{equation}\label{general_eigen}
\[\Gamma_j, \bar{A}_x x + \bar{A}_t t \] = \eta_j(x,t)\Gamma_j.
\end{equation}
The function $\eta_j$ encodes the \emph{dispersion relation}. 
From \eqref{psi_vacuum} we then have
\begin{equation}\label{psigpsi}
\bar{\Psi} g \bar{\Psi}^{-1} = \prod_{j=1}^{n}\bar{\Psi}\exp\(\Gamma_j\)
\bar{\Psi}^{-1}
= \prod_{j=1}^{n}\exp\(\bar{\Psi}\Gamma_j\bar{\Psi}^{-1}\)
= \prod_{j=1}^{n}\exp\(e^{\eta_j}\Gamma_j\).
\end{equation}
The vertex operators satisfy the nilpotency property between the states, which
eliminates terms containing its powers, $\(\Gamma_j\)^m$ for $m\ge2$, and
truncates the exponential series in \eqref{psigpsi}. 
Therefore, the $\tau$-functions \eqref{tau} assume the following form
\begin{equation}\label{tau_g}
\tau_{ab} = \bra{\l_a} \prod_{j=1}^n\(1+e^{\eta_j(x,t)}\Gamma_j\) \ket{\l_b}.
\end{equation}

Let us now apply these general concepts to the KN hierarchy. Consider
the following two vertex operators
\begin{equation}\label{vertices}
%\begin{align}
\Gamma_j \equiv \sum_{n=-\infty}^{\infty}\k_j^{-n}\Ea{n-1}, 
%\label{vert1} \\
\qquad
\Gamma'_j \equiv \sum_{n=-\infty}^{\infty}\k_j^{-n}\Eb{n},
%\label{vert2}
%\end{align}
\end{equation}
which satisfy the following eigenvalue commutation relations with
the vacuum \eqref{vacuum}
\begin{equation}\label{eigen}
\[ \Gamma_j, \H{m} \] = -2\k_j^{m}\Gamma_j, \qquad
\[ \Gamma'_j, \H{m} \] = 2\k_j^{m}\Gamma'_j.
\end{equation}
The dispersion relations of the KN hierarchy are, therefore, given by
\begin{equation}\label{dispersion}
\eta_j = -2\k_jx - 2\k_j^N t, \qquad
\eta'_j = 2\k_jx + 2\k_j^N t,
\end{equation}
where $N$ is the integer labelling the respective
flow, i.e. $N=2,3,\dotsc$ for the positive flows or $N=-1,-2,\dotsc$
for the negative flows. 
Thus \eqref{dispersion} explains the change in
the dispersion relation between models \eqref{mplr} and \eqref{dnlse} 
\cite{matsuno_fl}.

\subsection{Two vertices solution}
\begin{figure}
\begin{center}
\includegraphics{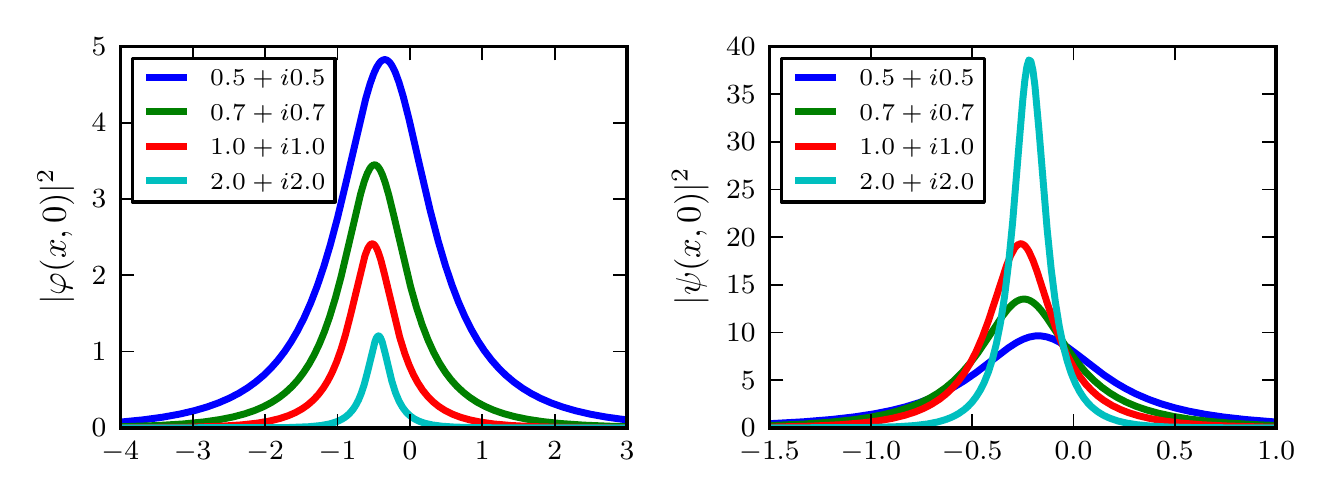}
\caption{Graphs of \eqref{mplr_square} and \eqref{dnlse_square}, 
corresponding to one-soliton solution of \eqref{mplr} and \eqref{dnlse}
(with the first sign), respectively, for different values of 
$\k = \k_R + i\k_I$. Note the peculiar feature of the first graph, 
where the soliton becomes wider when it gets higher,
contrary to the second graph that exhibits the usual soliton behaviour.}
\label{fig:1}
\end{center}
\end{figure}

Using a single vertex operator in \eqref{tau_g} we get a non interesting 
simple exponential solution with one of the fields vanishing, which 
corresponds to a linearization of the equations of motion. 
The first nontrivial solution is obtained with two vertices in the form
\begin{equation}
g=\exp\(\Gamma_1\)\exp\(\Gamma'_2\).
\end{equation}
Then, from \eqref{tau_g} we obtain
\begin{equation}\label{tau_2vert}
\tau_{ab} = \bra{\l_a} 1 + \Gamma_1 e^{\eta_1} + \Gamma'_2 e^{\eta'_2} +
\Gamma_1\Gamma'_2 e^{\eta_1+\eta'_2} \ket{\l_b}.
\end{equation}
The matrix elements are calculated in appendix \ref{sec:matrix1}, leading
to the following $\tau$-functions
\begin{equation}\label{tau_2vertices}
\begin{aligned}
\tau_{00} &= 1+\dfrac{\k_2}{\(\k_1-\k_2\)^2}e^{\eta_1+\eta'_2}, & \qquad
\tau_{02} &= \dfrac{1}{\k_2}e^{\eta'_2}, & \qquad
\tau_{13} &= \dfrac{1}{\k_1}e^{\eta_1}, \\
\tau_{11} &= 1+\dfrac{\k_1}{\(\k_1-\k_2\)^2}e^{\eta_1+\eta'_2}, &
\tau_{20} &= e^{\eta_1}, &
\tau_{31} &= e^{\eta'_2}.
\end{aligned}
\end{equation}
Replacing \eqref{tau_2vertices} in relations 
\eqref{sol_mpl}--\eqref{sol_thirr} we can obtain explicit one-soliton 
solutions for the all the previous models considered in this paper. 
Let us also recall that we must pick the right dispersion 
relation \eqref{dispersion} for the corresponding flow $N$.
For instance, from \eqref{sol_mpl} we have a
solution of \eqref{mpl}, that under the reduction
$x\to\tfrac{i}{2}x$, $t\to-\tfrac{i}{2}t$ and $\k_2=\k_1^*=\k$, which
is compatible with the choice $\phi=\rho^*=\varphi$, yields a solution
of \eqref{mplr} with the \emph{minus sign}. Still writing $\k = \k_R + i\k_I$
we have found
\begin{equation}\label{mplr_square}
|\vphi|^2 = \dfrac{\(\k_R^2+\k_I^2\)^{-1}e^{\theta}}{
1-\dfrac{\k_R}{2\k_I^2}e^{\theta}+\dfrac{\k_R^2+\k_I^2}{16\k_I^4}
e^{2\theta}}, \qquad
\theta = -2\k_I\(x - \dfrac{1}{\k_R^2+\k_I^2}t\).
\end{equation}
Using \eqref{sol_pos} we have a solution of \eqref{dnls}, 
that with the appropriate reduction yields a solution of
\eqref{dnlse} with the \emph{plus sign}, whose square modulus reads
\begin{equation}\label{dnlse_square}
|\psi|^2 = \dfrac{4e^\theta}{1-\dfrac{\k_R}{2\k_I^2}e^{\theta}+
\dfrac{\k_R^2+\k_I^2}{16\k_I^4}e^{2\theta}}, \qquad
\theta = -4\k_I\(x+4\k_Rt\).
\end{equation}
Both functions \eqref{mplr_square} and \eqref{dnlse_square} are, respectively,
plotted in figure \ref{fig:1}. The solution \eqref{mplr_square} has an
unusual behaviour, as can be seen from its graph. The soliton
gets wider as its height and velocity increases. 
This behaviour does not occur for
\eqref{dnlse_square} that shows the usual solitonic profile.

From \eqref{sol_thirr} we have the following one-soliton solution for
the Thirring model \eqref{thirr1}
\begin{equation}\label{onesol_thirr}
u = \dfrac{i}{\sqrt{2g}}\dfrac{e^{\eta^*}}{
1+ \dfrac{\k^*}{\(\k-\k^*\)^2}e^{\eta+\eta^*}}, \qquad
v = \dfrac{im}{\sqrt{2g}} 
\dfrac{\(\k^*\)^{-1}e^{\eta^*}}{1+\dfrac{\k}{\(\k-\k^*\)^2}
e^{\eta+\eta^*}},
\end{equation}
where $\eta = -i\(\k x - m^2 \k^{-1}t\)$. Taking the square modulus of these
solutions we obtain exactly the behaviour of \eqref{mplr_square} for $v$
and the behaviour of \eqref{dnlse_square} for $u$, which are sketched in
figure \ref{fig:1}. The formulas are almost identical.

\subsection{Four vertices solution}
\begin{figure}
\begin{center}
\includegraphics{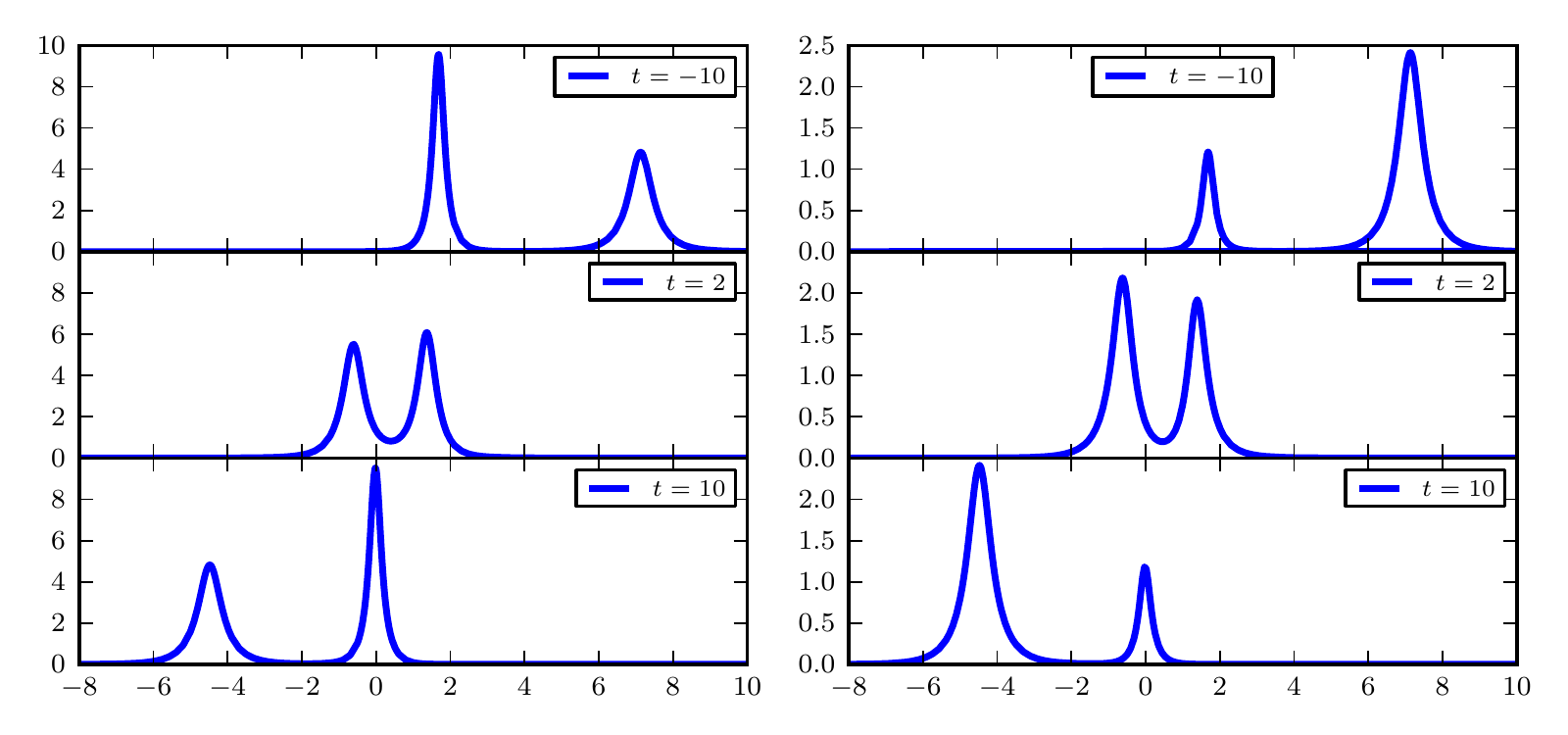}
\caption{Graphs of solutions \eqref{sol_thirr} involving four vertices.
In the left column we have ploted the time evolution of $|u|^2$ 
and in the right column the time evolution of $|v|^2$. We have
set the parameters as $m=1$, $g=\tfrac{1}{2}$, $\k_2=\k_1^*=\k=1+i$
and $\k_4=\k_3^*=\zeta=2+2i$. The waves travel from right to left.
Note that for $|u|^2$ the small soliton is faster than the largest one,
while for $|v|^2$ the higher soliton is faster.}
\label{fig:2}
\end{center}
\end{figure}

Let us consider a more complex solution by choosing the group
element with four vertices in the form
\begin{equation}
g = \exp\(\G_1\)\exp\(\G'_2\)\exp\(\G_3\)\exp\(\G'_4\).
\end{equation}
We then calculate the $\tau$-functions analogously to \eqref{tau_2vert}. 
After calculating the matrix elements, which are presented
in appendix \ref{sec:matrix1}, we obtain 
\begin{subequations}\label{tau_4vertices}
\begin{align}
\tau_{00} &= \bra{\l_0} 1+
\G_1\G'_2 e^{\eta_1+\eta'_2}+
\G_1\G'_4 e^{\eta_1+\eta'_4}+
\G'_2\G_3 e^{\eta'_2+\eta_3}+
\G_3\G'_4 e^{\eta_3+\eta'_4} \nn \\
&\qquad \qquad
+\G_1\G'_2\G_3\G'_4 e^{\eta_1+\eta'_2+\eta_3+\eta'_4}\ket{\l_0}, \\
%%%
\tau_{11} &= \bra{\l_1} 1+
\G_1\G'_2 e^{\eta_1+\eta'_2}+
\G_1\G'_4 e^{\eta_1+\eta'_4}+
\G'_2\G_3 e^{\eta'_2+\eta_3}+
\G_3\G'_4 e^{\eta_3+\eta'_4} \nn \\
&\qquad \qquad
+\G_1\G'_2\G_3\G'_4 e^{\eta_1+\eta'_2+\eta_3+\eta'_4}\ket{\l_1}, \\
%%%
\tau_{02} &= \bra{\l_0} \G'_2 e^{\eta'_2}+\G'_4 e^{\eta'_4}
+\G'_2\G_3\G'_4 e^{\eta'_2+\eta_3+\eta'_4}
+\G_1\G'_2\G'_4 e^{\eta_1+\eta'_2+\eta'_4} \ket{\l_2}, \\
%%%
\tau_{13} &= \bra{\l_1} \G_1 e^{\eta_1}+\G_3 e^{\eta_3}
+\G_1\G'_2\G_3 e^{\eta_1+\eta'_2+\eta_3}
+\G_1\G_3\G'_4 e^{\eta_1+\eta_3+\eta'_4} \ket{\l_3}, \\
%%%
\tau_{20} &= \bra{\l_2} \G_1 e^{\eta_1}+\G_3 e^{\eta_3}
+\G_1\G'_2\G_3 e^{\eta_1+\eta'_2+\eta_3}
+\G_1\G_3\G'_4 e^{\eta_1+\eta_3+\eta'_4} \ket{\l_0}, \\
%%%
\tau_{31} &= \bra{\l_3} \G'_2 e^{\eta'_2}+\G'_4 e^{\eta'_4}
+\G'_2\G_3\G'_4 e^{\eta'_2+\eta_3+\eta'_4}
+\G_1\G'_2\G'_4 e^{\eta_1+\eta'_2+\eta'_4} \ket{\l_1}.
\end{align}
\end{subequations}
Considering special transformations,
for instance, $\k_1=\k_2^*=\k$ and $\k_3=\k_4^*=\zeta$ one obtains
a two-soliton solution in the same way as 
\eqref{mplr_square}--\eqref{onesol_thirr}.
We will not write down further explicit
formulas but in figure \ref{fig:2} we show a graph of the two-soliton 
solution of the Thirring model \eqref{thirr1} obtained in this way
from \eqref{sol_thirr}. 

The one- and two-soliton solutions, given by \eqref{tau_2vertices} and 
\eqref{tau_4vertices}, respectively, 
were explicitly checked against the all the previous models mentioned
in this paper, with the aid of symbolic computation. We also want
to stress that in spite of the technical difficulty, the procedure
to compute the $\tau$-functions with $n$ vertices is well defined and
systematic.

\section{Equivalent construction}
\label{sec:equivalency}

The standard construction of the KN hierarchy is given by the
following Lax operator, with the \emph{homogeneous gradation}
\cite{kaup_newell,gerdjikov_kulish,gerdjikov,mikhailov_thirring}
\begin{equation}\label{standard_kn}
A_x = \H{2} + q\Ea{1} + r\Eb{1} = 
\begin{pmatrix} \l^2 & \l q \\ \l r & -\l^2 \end{pmatrix}, \qquad
Q = \hd = \l\dfrac{d}{d\l}.
\end{equation}
Given an affine Lie algebra $\alie$ with generators $T_a^n$, the 
\emph{conjugation} by a group element $h$,  
i.e. $T_a^n \mapsto hT_a^nh^{-1}$, is an \emph{automorphism}
since it preserves the commutator \eqref{kac_commutator}. Consider
the algebra $\hat{A}_1$ and let us define the following group element
\begin{equation}\label{automorphism}
h \equiv \exp\left\{-\tfrac{1}{2}\ln\(\l\) \H{0}\right\},
\end{equation}
which yields the following mapping
\begin{equation}
\begin{split}
h\H{n}h^{-1} &= \H{n}, \\
h\hc h^{-1} &= \hc,
\end{split}\qquad
\begin{split}
h\Ea{n}h^{-1} &= \Ea{n-1}, \\
h\hd h^{-1} &= \hd + \tfrac{1}{2}\H{0},
\end{split}\qquad
\begin{split}
h\Eb{n}h^{-1} &= \Eb{n+1}.
\end{split}
\end{equation}
Thus \eqref{standard_kn} is mapped into
\begin{equation}\label{almost_mapping}
A_x\mapsto hA_xh^{-1} = \H{2} + q\Ea{0} +r\Eb{2}, \qquad
Q\mapsto hQh^{-1}=\tfrac{1}{2}\H{0}+\hd. 
\end{equation}
If we now introduce a \emph{new spectral parameter} $\zeta\equiv\l^2$, then 
$\zeta\tfrac{d}{d\zeta}=\tfrac{1}{2}\l\tfrac{d}{d\l}$ and by redefining
the operators in \eqref{almost_mapping} with respect to the spectral
parameter $\zeta$, we map \eqref{standard_kn} into another
algebraic construction with \emph{principal gradation}
\begin{equation}\label{equivalent_construction}
\begin{cases}
\begin{aligned}[t]
A_x &= \H{2}+q\Ea{1}+r\Eb{1} \\
&= \begin{pmatrix}\l^2 & \l q \\ \l r & -\l^2 \end{pmatrix} \\
Q &= \hd
\end{aligned}
\end{cases}
\begin{aligned}[c]
\xrightarrow[\zeta = \l^2]{h = e^{-1/2\ln\l\H{0}}}
\end{aligned}\quad
\begin{cases}
\begin{aligned}[t]
hA_xh^{-1} &= \H{1}+q\Ea{0}+r\Eb{1} \\
&= \begin{pmatrix}\zeta & q \\ \zeta r & -\zeta \end{pmatrix} \\
hQh^{-1} &= \tfrac{1}{2}\H{0} +2\hd
\end{aligned}
\end{cases}
\end{equation}
Therefore, we have demonstrated the equivalence of the standard
construction \eqref{standard_lax_kn} and the one proposed in this paper
\eqref{kn_potentials}.
We have introduced the construction \eqref{kn_potentials} because it is 
much simpler and natural to be treated under the dressing method 
than \eqref{standard_kn}, and it eliminates the spurious quadratic
power of the spectral parameter that is responsible for divergences
in some complex integrals appearing in the IST.

\paragraph{Comment.} It is possible to apply the method used in
this paper to the homogeneous construction \eqref{standard_lax_kn} and the
same relations \eqref{A0}--\eqref{sol_neg} arises, but the vacuum 
\eqref{psi_vacuum} now stays in the form 
\begin{equation}\label{new_vac}
\bar{\Psi} = e^{-\H{2}x}e^{-\H{2N}t}e^{-2xt\d_{N+1,0}\hc}.
\end{equation}
The vertex operators \eqref{vertices} still statisfy the eigenvalue relations
with this vacuum but are not uniformely graded according to the homogeneous
gradation. Moreover, to be able to obtain the \emph{correct} solution
we must redefine the spectral parameter in \eqref{new_vac} $\zeta\equiv\l^2$.
Therefore, the procedure does not occur in a natural way as in the case
of principal gradation and it is necessary to introduce some ingredients
by hand.

\section{Concluding remarks}
\label{sec:conclusions}

The KN hierarchy was obtained from a \emph{higher grading}
affine algebraic construction with the algebra
$\hat{A}_1$ and \emph{principal gradation}.
In fact, we have proposed a general construction  
that can generates novel integrable models if different affine Lie algebras  
are employed. The results of this paper should extend naturally to 
these cases.

The main models within the KN hierarchy were derived. The DNLSE-I 
\eqref{dnlse} arises from the second positive flow, while
the Mikhailov model \eqref{mplr} is obtained from the first negative 
flow. The gauge transformation \eqref{tilde_fields} 
connects the system \eqref{dnls} to \eqref{tilde_eqs}, which in particular
yields relations between the three kinds of DNLS 
equations, namely, \eqref{dnlse}, \eqref{dnls2} and \eqref{dnls3}. 
Furthermore, we have demonstrated a general relation between
the model \eqref{mplr} and the massive Thirring model 
\eqref{thirr1} through \eqref{new_fields}.

We developed the algebraic dressing method for the KN hierarchy and
several relations found previously in the literature emerges naturally.
For instance, the form of the gauge 
transformations linking the three DNLS equations and the precise connection 
between the solutions of \eqref{mplr} and those of \eqref{dnlse}. 
Moreover, the weight function introduced in the revised IST 
\cite{kaup_newell} also arises from the algebraic dressing method.
We stress that this method is general and systematic, and relies only
on the algebraic structure of the hierarchy.
The solutions of all models considered in this paper were expressed in terms 
of $\tau$-functions, which can be systematically calculated through a
vertex representation theory of the algebra.
We considered explicitly one- and two-soliton solutions. 
The solitons of the model \eqref{mplr}, given by \eqref{mplr_square}, 
possess an unusual behaviour were its width increases with its height, 
as shown in figure \ref{fig:1}.
%These results provide a unifying picture. Models that were previously 
%studied in a separate and unrelated way, were treated under the same 
%algebraic structure.

Finally, we demonstrated that our construction \eqref{kn_potentials} is 
conjugate related to the usual construction found in the literature
\eqref{standard_lax_kn}. However, the dressing procedure applied to our 
Lax pair \eqref{kn_potentials} is greatly simplified compared 
to \eqref{standard_lax_kn}.
Several works on VBC and also NVBC are based on the standard Lax pair
and the revised IST.
We conclude that these problems can be simplified  
if one considers our construction instead. We plan to illustrate this fact more
precisely regarding NVBC within the dressing formalism in a future 
opportunity.

\subsubsection*{Acknowledgments}
We thank CAPES, CNPq and Fapesp for financial support.
GSF thanks the support from CNPq under the 
``Ci\^ encia sem fronteiras'' program.

\appendix

\section{Algebraic concepts}
\label{sec:algebra}

Let $\lie$ be a finite dimensional Lie algebra, with commutator 
$\[ T_a, T_b \]$ for $T_a,T_b \in \lie$ and with a symmetric bilinear 
Killing form $\inner{T_a}{T_b}$. 
The \emph{infinite} dimensional loop-algebra is defined by
$\llie \equiv \lie\otimes\complex(\l,\l^{-1})$, i.e. 
$T_a \mapsto T_a\otimes\lambda^n\equiv T_a^n \in \llie$ for $n\in \integer$
and $\l$ is the so called complex \emph{spectral parameter}.
Let us introduce the central term $\hc$, that commutes with every other 
generator, and also the derivative operator 
$\hd \equiv \lambda \tfrac{d}{d\lambda}$. 
The Kac-Moody algebra is then defined by
$\alie \equiv \llie\oplus\complex\hc\oplus\complex\hd$ with commutator
\begin{equation}\label{kac_commutator}
\blb T_a^n, T_b^m \brb \equiv \blb T_a,T_b \brb \otimes \lambda^{n+m} + \hc
n\delta_{n+m,0}\inner{T_a}{T_b}.
\end{equation}
If we set $\hc = 0$ we have the commutator for the loop-algebra $\llie$.
For example, considering the algebra $A_1 = \blcurl E_\a, E_{-\a}, H \brcurl$, 
where $\blb E_{\a}, E_{-\a} \brb = H$ and 
$\blb H, E_{\pm\a}\brb = \pm 2 E_{\pm\a}$, we have
the Kac-Moody algebra $\hat{A}_1$ with generators
$\blcurl E_\a^{n}, \, E_{-\a}^{n}, \, H^n, \, \hc, \hd  \brcurl$ 
and commutation relations
\begin{equation}\label{commutators}
\begin{split}
\blb H^{n}, H^{m} \brb &= 2n\d_{n+m,0}\hc, \\
\blb H^{n}, E^{m}_{\pm\a} \brb &= \pm 2 E^{n+m}_{\pm\a},
\end{split}\qquad
\begin{split}
\blb E^n_{\a}, E^{m}_{-\a} \brb &= H^{n+m} + n\d_{n+m,0}\hc, \\
\blb \hd, T^{n} \brb &= nT^{n}, \\
\end{split}\qquad
\begin{split}
\blb \hc, T^{n} \brb = 0, 
\end{split}
\end{equation}
where $T^{n} \in \lcurl H^{n},\,E^{n}_{\a}, \, E^{n}_{-\a}\rcurl$.

We can introduce a grading operator $Q$, that splits the algebra into 
graded subspaces in the following way. For $T_a^{n} \in \alie$, if  
$\[ Q, T_a^{n} \] = m T_a^{n}$ for $m \in \integer$, then
$\alie = \bigoplus_{m\in\integer}\alie^{(m)}$ where
$\alie^{(m)}=\blcurl T_a^{n} \, | \, \[Q,T_a^n\]=mT_a^{n} \brcurl$.
In the case of $\alie=\hat{A}_1$ the grading
operator $Q\equiv \hd$ (homogeneous) induces a natural gradation,
$\alie^{(m)} = \lcurl H^m, E_{\a}^m, E_{-\a}^m \rcurl$. For
$Q\equiv \tfrac{1}{2}\H{0}+2\hd$ (principal) we have
$\alie^{(2m+1)} = \lcurl E_{\a}^m, E_{-\a}^{m+1}\rcurl$ and
$\alie^{(2m)} = \lcurl H^m \rcurl$.
The operators $\hc$ and $\hd$ have zero grade.

The highest weight states of the algebra is a set of states
satisfying $T_a^n\ket{\l_a}=0$, if $T_a^n$ have grade higher than zero.
Precisely for the case of $\hat{A}_1$, the highest
weight states are $\lcurl \ket{\l_0}, \ket{\l_1} \rcurl$ and obey 
the following actions
\begin{equation}
\begin{aligned}[c]
E_{\pm \a}^{n}\ket{\l_a} &= 0 \quad (n > 0), \\
H^{n}\ket{\l_a} &= 0 \quad (n > 0),
\end{aligned}\qquad
\begin{aligned}[c]
H^{0}\ket{\l_0} &= 0, \\
H^{0}\ket{\l_1} &= \ket{\l_1},
\end{aligned}\qquad
\begin{aligned}[c]
\Ea{0}\ket{\l_a} &= 0, \\
\hc\ket{\l_a} &= \ket{\l_a},
\end{aligned}
\end{equation}
where $a=0,1$.
The adjoint relations are $\(\H{n}\)^{\dagger}=\H{-n}$,
$\(E_{\pm\a}^{n}\)^{\dagger}=E_{\mp\a}^{-n}$ and 
$\hc^{\dagger}=\hc$.

A $2\times2$ matrix representation of 
the $\hat{A}_1$ generators can be given as
follows
\begin{equation}
\H{n} = \begin{pmatrix} \l^n & 0 \\ 0 & -\l^n \end{pmatrix}, \quad
\Ea{n} = \begin{pmatrix} 0 & \l^n \\ 0 & 0 \end{pmatrix}, \quad
\Eb{n} = \begin{pmatrix} 0 & 0 \\ \l^n & 0 \end{pmatrix}, \quad
\hc = \begin{pmatrix} 1 & 0 \\ 0 & 1\end{pmatrix}.
\end{equation}

\section{Matrix elements}
\label{sec:matrix1}

The relevant states are 
$\ket{\l_0}$, $\ket{\l_1}$, $\ket{\l_2} = \Ea{-1}\ket{\l_0}$
and $\ket{\l_3} = \Eb{0}\ket{\l_1}$. The vertices for the KN hierarchy are
\begin{equation}
\G_j = \sum_{n=-\infty}^{\infty}\k_j^{-n}\Ea{n-1}, \qquad
\G'_j = \sum_{n=-\infty}^{\infty}\k_j^{-n}\Eb{n},
\end{equation}
and satisfy the eigenvalue equations \eqref{eigen}.
We will write only the \emph{nonvanishing} matrix elements used in 
\eqref{tau_2vert} and \eqref{tau_4vertices}.
The \emph{nilpotency} property of the vertices reads
$\bra{\l_a} \(\Gamma_i\)^n \ket{\l_b} = 
\bra{\l_a} \(\Gamma'_i\)^n \ket{\l_b} = 0$ for $n\ge2$. In addition,
any matrix element having a power of a vertex vanishes, e.g.
$\bra{\l_a} \(\Gamma_i\)^2\Gamma'_j \ket{\l_b} = 0$. Another useful
result is that the number of $\Ea{n}$ and $\Eb{m}$ in a matrix element must
be balanced in pairs, e.g.  
$\bra{\l_a}\G_i\G'_j\ket{\l_a}\ne 0$, while
$\bra{\l_0}\G_i\G'_j\ket{\l_2}=\bra{\l_0}\G_i\G'_j\Ea{-1}\ket{\l_0}=0$.
Thus, the nonvanishing matrix elements relevant
for the solution with two vertices are
\begin{align}
\bra{\l_2}\Gamma_1\ket{\l_0} &= 1, \\
\bra{\l_1}\Gamma_1\ket{\l_3} &= \dfrac{1}{\k_1}, \\
\bra{\l_0}\Gamma'_2\ket{\l_2} &= \dfrac{1}{\k_2}, \\
\bra{\l_3}\Gamma'_2\ket{\l_1} &= 1, \\
\bra{\l_0}\Gamma_1\Gamma'_2\ket{\l_0} &= \dfrac{\k_2}{\(\k_1-\k_2\)^{2}}, \\
\bra{\l_1}\Gamma_1\Gamma'_2\ket{\l_1} &= \dfrac{\k_1}{\(\k_1-\k_2\)^{2}}.
\end{align}
Note also that $\bra{\l_a}\G'_2\G_1\ket{\l_a}=\bra{\l_a}\G_1\G'_2\ket{\l_a}$.
Besides these elements, the nonvanishing elements for the 
solution with four vertices are
\begin{align}
\bra{\l_0}\G'_2\G_3\G'_4\ket{\l_2} &= 
\dfrac{\k_3^2\(\k_2-\k_4\)^2}{\k_2\k_4\(\k_2-\k_3\)^2\(\k_3-\k_4\)^2}, \\
%%%
\bra{\l_0}\G_1\G'_2\G'_4\ket{\l_2} &= 
\dfrac{\k_1^2\(\k_2-\k_4\)^2}{\k_2\k_4\(\k_1-\k_2\)^2\(\k_1-\k_4\)^2}, \\
%%%
\bra{\l_1}\G_1\G'_2\G_3\ket{\l_3} &= 
\dfrac{\k_2^2\(\k_1-\k_3\)^2}{\k_1\k_3\(\k_1-\k_2\)^2\(\k_2-\k_3\)^2}, \\
%%%
\bra{\l_1}\G_1\G_3\G'_4\ket{\l_3} &= 
\dfrac{\k_4^2\(\k_1-\k_3\)^2}{\k_1\k_3\(\k_1-\k_4\)^2\(\k_3-\k_4\)^2}, \\
%%%
\bra{\l_2}\G_1\G'_2\G_3\ket{\l_0} &= 
\dfrac{\k_2\(\k_1-\k_3\)^2}{\(\k_1-\k_2\)^2\(\k_2-\k_3\)^2}, \\
%%%
\bra{\l_2}\G_1\G_3\G'_4\ket{\l_0} &= 
\dfrac{\k_4\(\k_1-\k_3\)^2}{\(\k_1-\k_4\)^2\(\k_3-\k_4\)^2}, \\
%%%
\bra{\l_3}\G'_2\G_3\G'_4\ket{\l_1} &= 
\dfrac{\k_3\(\k_2-\k_4\)^2}{\(\k_2-\k_3\)^2\(\k_3-\k_4\)^2}, \\
%%%
\bra{\l_3}\G_1\G'_2\G'_4\ket{\l_1} &= 
\dfrac{\k_1\(\k_2-\k_4\)^2}{\(\k_1-\k_2\)^2\(\k_1-\k_4\)^2}, \\
%%%
\bra{\l_0}\G_1\G'_2\G_3\G'_4\ket{\l_0} &= 
\dfrac{\k_2\k_4\(\k_1-\k_3\)^2\(k_2-\k_4\)^2}{
\(\k_1-\k_2\)^2\(\k_1-\k_4\)^2\(\k_2-\k_3\)^2\(k_3-\k_4\)^2}, \\
%%%
\bra{\l_1}\G_1\G'_2\G_3\G'_4\ket{\l_1} &= 
\dfrac{\k_1\k_3\(\k_1-\k_3\)^2\(k_2-\k_4\)^2}{
\(\k_1-\k_2\)^2\(\k_1-\k_4\)^2\(\k_2-\k_3\)^2\(k_3-\k_4\)^2}. 
\end{align}

%\section*{References}

\bibliographystyle{JHEP}
\bibliography{biblio}

\end{document}

%% file: macros.tex
\def\nn{\nonumber}
\def\nnn{\nonumber\\}
\def\pa{\partial}

\newcommand{\be}{\begin{equation}}
\newcommand{\ee}{\end{equation}}

\def\({\left(}
\def\){\right)}

\renewcommand{\[}{\left[}
\renewcommand{\]}{\right]}
\def\blb{\bigl[}
\def\brb{\bigr]}
\def\lcurl{\left\{}
\def\rcurl{\right\}}
\def\blcurl{\big\{}
\def\brcurl{\big\}}

\def\a{\alpha}

\def\G{\Gamma}
\def\d{\delta}

\def\k{\kappa}
\def\l{\lambda}

\def\vphi{\varphi}

\def\bvphi{\bm{\vphi}}

\def\tpsi{\widetilde{\psi}}

\def\Thp{\Theta_{+}}
\def\Thm{\Theta_{-}}
\def\Thpm{\Theta_{\pm}}

\def\lie{\mathcal{G}}
\def\alie{{\widehat\lie}}
\def\llie{\widetilde\lie}
\def\cm{{\cal M}}
\def\ck{{\cal K}}

\def\H#1{H^{#1}}
\def\Ea#1{E_{\a}^{#1}}
\def\Eb#1{E_{-\a}^{#1}}

\def\D#1{D^{(#1)}}
\def\E#1{E^{(#1)}}
\def\F#1{F^{(#1)}}
\def\hc{\hat{c}}
\def\hd{\hat{d}}

\def\complex{\mathbb{C}}
\def\integer{\mathbb{Z}}

\def\bra#1{\langle#1|}
\def\ket#1{|#1\rangle}

\def\inner#1#2{\langle#1|#2\rangle}

\def\tq{\tilde{q}}
\def\tr{\tilde{r}}
\def\tpsi{\tilde{\psi}}
\def\A#1{\mathcal{A}^{(#1)}}
\def\B#1{\mathcal{B}^{(#1)}}